\documentclass{mn2e} 

\usepackage{amsfonts}
\usepackage{amsmath}
\usepackage{graphicx}

\title[Central and satellite colors in galaxy groups]
      {Central and Satellite Colors in Galaxy Groups: A Comparison of the Halo Model and SDSS Group Catalogs} 
\author[R. A. Skibba]
 {Ramin A. Skibba$^1$
  \thanks{E-mail:  skibba@mpia.de}\\
  $^{1}$Max-Planck-Institute for Astronomy, K\"{o}nigstuhl 17,
	D-69117 Heidelberg, Germany}

\begin{document}

\pagerange{\pageref{firstpage}--\pageref{lastpage}}

\maketitle
\label{firstpage}

\begin{abstract}
Current analytic and semi-analytic dark matter halo models distinguish between
the central galaxy in a halo and the satellite galaxies in halo substructures.
It is expected that galaxy properties are correlated with host
halo mass, and that central galaxies tend to be the most luminous, massive, and 
reddest galaxies in halos while the satellites around them are fainter and bluer.
Using a recent halo-model description of the color dependence 
of galaxy clustering (Skibba \& Sheth 2008), we investigate the colors of
central and satellite galaxies predicted by the model and compare them to
those of two galaxy group catalogs constructed from the Sloan Digital
Sky Survey (Yang et al. 2007, Berlind et al. 2006a).
In the model, the environmental dependence of galaxy color is determined by
that of halo mass, and the predicted color mark correlations were
shown to be consistent with SDSS measurements.
The model assumes that satellites tend to follow a color-magnitude sequence
that approaches the red sequence at bright luminosities; the model's success
suggests that 
bright satellites tend to be `red and dead' while the star
formation in fainter ones is in the process of being quenched.
In both the model and the SDSS group catalogs,
we find that at fixed luminosity or stellar mass,
central galaxies tend to be bluer than satellites.
In contrast, at fixed group richness or halo mass,
central galaxies tend to be redder than satellites, and 
galaxy colors become redder with increasing mass.
We also compare the central and satellite galaxy color distributions,
as a function of luminosity and as a function of richness,
in the model and in the two group catalogs.
Except for faint galaxies and small groups, the model and both 
group catalogs are in very good agreement.
\end{abstract}


\section{Introduction}\label{intro}

In standard $\Lambda$CDM cosmological models, cold dark matter halos form from 
the gravitational collapse of dark matter particles, and they assemble 
hierarchically, such that smaller halos merge to form larger and more massive
halos.
According to the current paradigm of galaxy formation, galaxies form within 
halos, due to the cooling of hot gas.
Halos and galaxies evolve simultaneously, and the evolution of a galaxy is 
affected by its host halo.   If the halo is accreted by a larger halo, the
galaxy will be affected by it as well, and may interact or merge with the 
galaxies within the new host halo.
Such `satellite' galaxies in halo substructures no longer accrete hot gas,
which instead is only accreted by the `central' galaxy in the halo.
The central galaxy consequently continues to grow, while other massive 
galaxies may merge into it, and therefore it is expected to be the most 
luminous and most massive galaxy in the halo.

For these reasons, current analytic and semi-analytic models distinguish 
between central and satellite galaxies, which at a given time are at different 
stages of evolution, or may have evolved differently.
As galaxies evolve they transform from star-forming late-type galaxies into 
massive bulge-dominated galaxies with little or no ongoing star formation.
It is thought that central galaxies undergo such a transformation by experiencing
a major merger followed by AGN feedback preventing additional gas cooling and star formation.
Satellite galaxies may have their star formation suppressed or `quenched' by a
number of other processes, such as ram-pressure stripping of the cold gas reservoir, 
`harassment' by other satellites, and `strangulation' following the stripping 
of the hot gas reservoir, the latter of which appears to be the dominant process
(\textit{e.g.}, Weinmann et al. 2006, van den Bosch et al. 2008a). 

Galaxies in relatively dense environments tend to reside in groups and clusters
hosted by massive halos.  Recent analyses with galaxy group catalogs have 
argued that many of these galaxies are very red with very low star formation 
rates, in contrast with galaxies in low-mass halos in less dense environments, 
many of which are still quite blue with significant star formation
(\textit{e.g.}, Weinmann et al. 2006, Berlind et al. 2006b).
Measurements of the environmental dependence of galaxy color have found trends 
that are qualitatively consistent with these claims (\textit{e.g.}, 
Zehavi et al. 2005, Blanton et al. 2005a, Tinker et al. 2007, Coil et al. 2008).
In order to better understand galaxy and halo evolution, more models are needed 
that can explain the environmental dependence of color, and more measurements 
of correlations between color and environment are needed to better constrain 
such models.
Skibba \& Sheth (2008) have taken a step in this direction: 
they developed and tested a halo model of the color dependence of 
galaxy clustering in the Sloan Digital Sky Survey (SDSS).
Their model successfully explains the correlation between color and environment,
quantified by the color mark correlation function, while assuming that all
environmental correlations are due to those of halo mass.
They distinguish between central and satellite galaxies, whose properties
are assumed to be determined by host halo mass.
The purpose of this paper is to further investigate these central and satellite 
galaxy colors, and in particular to compare the predictions of the model with 
measurements from recent galaxy group catalogs (Yang et al. 2007, Berlind et al. 2006a).

This paper is organized as follows.  In the next two sections, we briefly 
introduce the color mark model and the galaxy group catalogs.
In Section~\ref{groupcatcompare}, we compare the 
satellite color-magnitude sequence of the model to that of the Yang et al. catalog, 
and we compare the central and satellite colors of the model and both group 
catalogs as a function of group richness, which is a useful proxy for halo mass.
We summarize our results in Section~\ref{discuss}.

\section{Halo model of galaxy colors}\label{model} 

Our halo model of the color dependence of galaxy clustering is described in 
(Skibba \& Sheth 2008; hereafter SS08), 
and we refer the reader to this paper for details.

Briefly, our model is based on the model of luminosity dependent clustering of 
Skibba et al. (2006), which explained the observed environmental dependence of 
luminosity by applying the luminosity-dependent halo occupation distribution (HOD)
that was constrained by the observed luminosity-dependent correlation functions and 
galaxy number densities in the SDSS (Zehavi et al. 2005, Zheng et al. 2007).
The model of galaxy colors
in SS08 added constraints from the bimodal distribution of $g-r$ colors of SDSS 
galaxies as a function of $r$-band luminosity.  We made two assumptions: (i) that the 
bimodality of the color distribution at fixed luminosity is independent of
halo mass, and (ii) that satellite galaxies tend to follow a particular 
sequence in the color-magnitude diagram, one that approaches the red sequence
with increasing luminosity:
\begin{equation}
 \langle c|L\rangle_\mathrm{sat} = 
   \langle g-r|M_r\rangle_\mathrm{sat} \,=\, 0.83 - 0.08\,(M_r+20)
 \label{Csatseq}
\end{equation}

These observational constraints and additional assumptions allowed SS08 to
model the central and satellite galaxy color `marks' as a function of halo
mass, $\langle c|M\rangle_\mathrm{cen}$ and $\langle c|M\rangle_\mathrm{sat}$.
%
SS08 used the central and satellite galaxy marks to model color mark 
correlation functions, in which all correlations between color and environment
are due to those between halo mass and environment.
The modeled mark correlation functions were in very good agreement with their
measurements with volume-limited SDSS catalogs, reproducing the 
observed correlations between galaxy color and environment on scales of
$100h^{-1}\,\mathrm{kpc} < r_p < 30h^{-1}\,\mathrm{Mpc}$.

The two-point mark correlation function is simply the ratio $(1+W(r))/(1+\xi(r))$,
where $\xi(r)$ is the traditional two-point correlation function and $W(r)$ is
the same sum over galaxy pairs separated by $r$, but with each member of the pair
weighted by the ratio of its mark to the mean mark.
In practice, we measure the \textit{projected} clustering of galaxies, and
so we use the following analogous statistic for both the measurements and the models,
the marked projected correlation function:
\begin{equation}
  M_p(r_p)\,=\, \frac{1\,+\,W_p(r_p)/r_p}{1\,+\,w_p(r_p)/r_p} \, .
 \label{MCF}
\end{equation}
where the projected two-point correlation function is
\begin{equation}
  w_p(r_p)\,=\,2\, \int_0^\infty {\mathrm d}\pi\,\xi({r_p},\pi)\, 
            = \,2\, \int_{r_p}^\infty \,{\mathrm d}r\,
                         \frac{r\,\xi(r)}{\sqrt{r^2-{r_p}^2}},
\end{equation}
If mark correlations are consistent with unity, then the mark is not correlated
with the environment at that scale; if the mark correlations are above unity, which
is the case for galaxy luminosity and color (Skibba et al. 2006; SS08), then
higher values of the mark tend to be located in denser environments at that scale.

\begin{figure}
 \includegraphics[width=\hsize]{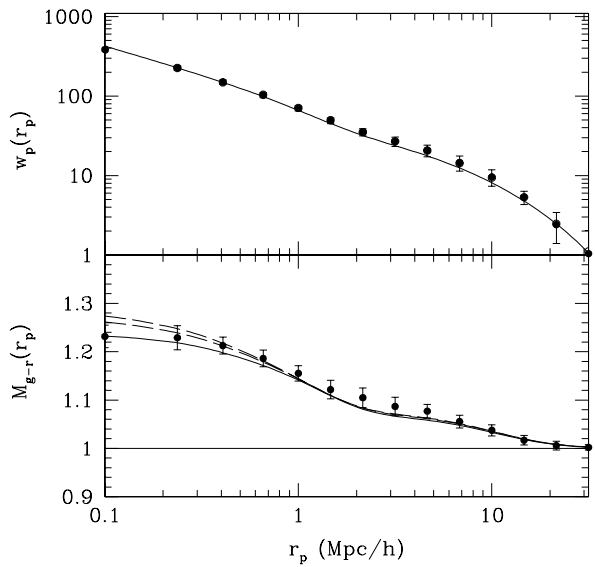} 
 \caption{Projected two-point correlation function and $g-r$ color mark
          correlation function for $M_r<-19.5$.
          Points show SDSS measurement for Petrosian colors,
          and solid curves show the fiducial model, both taken from Skibba \& Sheth (2008).
          Dashed curves show the range of color mark correlations predicted by the model
          if color gradients in groups and clusters are included (Hansen et al. 2007).}
 \label{MCFexample}
\end{figure}

As an example, we show the $g-r$ color mark correlation function for $M_r<-19.5$
in Figure~\ref{MCFexample}, reproduced from SS08 (their Figure 6).
The solid curve shows the halo model's prediction, which is in
agreement with the measurements of the color mark correlations
in the SDSS, using Petrosian colors.
Had we instead used $L_r/L_g$ as the mark, the resulting mark correlations
would be quantitatively stronger, but with larger uncertainties---the SDSS
measurements would have similar statistical significance and constraining
power as the $g-r$ mark measurements (see Skibba et al. 2006).

A few assumptions were made in the model that are worth discussing.
Firstly, it was assumed that the central galaxies lie at the center of their 
host dark matter halos, as is commonly assumed in HOD and conditional luminosity 
function (CLF) studies.  However, central galaxies are often offset from the center of 
the potential well (van den Bosch et al. 2005), and this offset appears to 
weakly dependent on galaxy color (Skibba et al. 2008b, in prep.).
Secondly, it was assumed that satellite galaxies follow the dark matter profile, 
while satellite galaxy number density profiles have been found to be less concentrated 
(\textit{e.g.}, Hansen et al. 2005, Yang et al. 2005b).
These two effects are both too weak to significantly affect the marked galaxy
clustering on scales of $r_p>100\,\mathrm{kpc}/h$, however, and if the effects 
were stronger, then there would have been a discrepancy with the observed 
unmarked correlation function $w_p(r_p)$ as well, which was not the case.

Thirdly, we also assumed that there are not mark gradients within halos, that is, that
the colors of galaxies are independent of their distance from the halo center.
In contrast, Hansen et al. (2007) have shown that the $g-r$ colors of galaxies 
are $\sim15-20\%$ redder in the inner regions of groups and clusters compared 
to the cluster outskirts (see their Figure 10).  
Van den Bosch et al. (2008b) have shown a similar fractional increase in
galaxy colors with decreasing halo-centric radius, down to halo masses of 
$10^{12}\,h^{-1}\,M_\odot$. 
These color gradients appear 
to be stronger than luminosity gradients in groups and clusters (Hansen et al. 
2005, Mart\'{i}nez \& Muriel 2006, Weinmann et al. 2006).
Such a positional dependence of galaxy colors would only affect the color
mark correlations while leaving the unmarked correlation function unchanged.
In order to model position-dependent color marks in mark clustering 
statistics, the density profile of satellite galaxies must be replaced by a 
weighted profile.  The calculation is done in Fourier space (see Sheth 2005, 
and Appendix B of SS08), and we replace $u_\mathrm{gal}(k|M)$, which is the 
Fourier transform of $\rho_\mathrm{gal}(r|M)$, by a weighted number density profile
\begin{equation}
   w(k|M) \,=\,
    \frac{\int \mathrm{d}r'\,4\pi r'^2\,c_\mathrm{gal}(r'|M)\,\rho_\mathrm{gal}(r'|M)\,\mathrm{sin}(kr')/kr'}
	 {\int \mathrm{d}r'\,4\pi r'^2\,c_\mathrm{gal}(r'|M)\,\rho_\mathrm{gal}(r'|M)} ,
\end{equation}
where $r'\equiv r/r_\mathrm{vir}(M)$ and $c_\mathrm{gal}(r'|M)$ quantifies
the dependence of galaxy color on halo-centric radius, which we assume to
be independent of mass because it is independent of cluster richness 
(Hansen et al. 2007; \textit{cf}., van den Bosch et al. 2008b).
The minimum and maximum amount of position-dependence of galaxy colors
result in slightly stronger color mark correlations at small scales 
(lower and upper dashed curves in lower panel of Figure~\ref{MCFexample}).
or in other words, color gradients in halos imply a slightly stronger
environmental dependence of galaxy color in halo environments ($r_p<1\,\mathrm{Mpc}/h$).
Note that color gradients in clusters are much stronger at $z>0.5$ (Loh et al. 2008)
and would have a stronger effect on color mark clustering at such redshifts.

Finally, the analytic halo-model description of the color dependence of galaxy clustering
uses the mean of the halo occupation distribution and the mean colors of central and
satellite galaxies.  We are assuming that at fixed halo mass, the halo occupation
distribution and color distributions are approximately independent of the environment---an
assumption which is justified by some recent results (Blanton \& Berlind 2007, 
van den Bosch et al. 2008b, Skibba \& Sheth 2008).
To explore the central and satellite galaxy color distributions predicted by the model,
we construct mock galaxy catalogs.  The mock catalogs are observationally constrained
by the halo occupation distribution as a function of luminosity, determined 
from galaxy clustering measurements in the SDSS (Zheng et al. 2007),
and the bimodal color distribution as a function of luminosity in the SDSS,
fit as the sum of two Gaussian distributions, which we refer to as the `red sequence' and the `blue sequence'.
In effect, since central and satellite galaxies have a range of colors 
at fixed luminosity, in the model satellite colors are drawn from the red sequence
with some luminosity dependent probability and are otherwise drawn from the blue sequence;
central galaxy colors are drawn with a similar procedure, but with a
different luminosity-dependent probability.
The details of the algorithm are described in Skibba et al. (2006) and SS08.
SS08 showed that these mock catalogs, which include not only the means of the
galaxy colors but their scatter as well, yield color mark correlation functions that
are consistent with the analytic model and with SDSS measurements.

The analysis in Section~\ref{groupcatcompare}, in which we examine the
colors of central and satellite galaxies, will constitute a test of 
this model, and in particular of the model's two assumptions 
described at the beginning of this section.
We will compare the model's predictions
of central and satellite galaxy colors to those of galaxy group
catalogs, described in the following section.

\section{Data: SDSS galaxy group catalogs}\label{cats} 

We will first compare our model to the Yang et al. (2007; hereafter Y07) group catalog, 
which was constructed by applying the halo-based group finder of Yang et al. (2005a) to
the New York University Value-Added Galaxy Catalog (NYU-VAGC; Blanton et al. 2005b),
which is based on the Sloan Digital Sky Survey (SDSS, York et al. 2000)
Data Release 4 (Adelman-McCarthy et al. 2006).

The group finder uses halo properties as a function of the total luminosity or
stellar mass of groups, and it also identifies `groups' with only a single member.
We used only those galaxies with spectroscopic redshifts from the SDSS or
with redshifts taken from other surveys (their `sample II')---that is, we
excluded fiber-collided galaxies that were not assigned fibers.
Including such galaxies does not affect the mean colors of central and satellite 
galaxies (shown in Section~\ref{groupcatcompare}), although this is not the case
for the mean central galaxy luminosities (see appendix of Skibba et al. 2007).
Y07 have used mock catalogs to account for the effects of the survey edges when 
estimating group halo masses (see their paper for details).
We only compare to volume-limited catalogs constructed from the sample II group
catalog in this work, and we have accounted for groups that overlap the 
redshift limits.
In each group, we identify the `central' galaxy as the most luminous in the 
$r$ band, and the remaining galaxies brighter than the luminosity threshold 
are the `satellites'.  Labeling the most massive galaxy in a group, rather 
than the brightest, as the central one does not affect our results.

We will also compare to the central and satellite galaxy colors of a
volume-limited group catalog of Berlind et al. (2006a; hereafter B06), which is
drawn from the SDSS large-scale structure sample \texttt{sample14} from the NYU-VAGC;
the sample is a subsample of SDSS DR4 (Adelman-McCarthy et al. 2006).
Fiber-collided galaxies are included, and each collided galaxy was given the
redshift of its nearest neighbor, which was shown to be an adequate correction
at least for groups with ten or more members.
To account for the survey edges, B06 excluded all groups whose centers lie less than 500 kpc from an edge in the tangential direction or less than 500 km $\mathrm{s}^{-1}$ from an 
edge in the radial direction.
Groups were identified using a halo-based friends-of-friends algorithm,
which used halo occupation distribution models and the group multiplicity function.
The group-finding algorithm only used the galaxy positions in redshift-space, as 
opposed to the algorithm used by Y07.

In the following analysis, the magnitudes and colors of galaxies in the two group
catalogs are Petrosian, and have been $k$-corrected and evolution corrected to $z=0.1$.

\section{Comparison of the Model and Group Catalogs}\label{groupcatcompare}

The main purpose of this paper is to analyze the colors of central
and satellite galaxies in groups and clusters.
We compare predictions of the color mark model of Skibba \& Sheth (2008)
to measurements from the Yang et al. (2007) and Berlind et al. (2006) galaxy group catalogs.
This analysis also constitutes a test of the model, which, as described in
Section~\ref{model}, made two assumptions in addition to those necessary to
model the luminosity dependence of galaxy clustering:
(i) that the bimodal color distribution at fixed luminosity is independent 
of halo mass, and 
(ii) that satellite galaxies tend to follow a particular color-magnitude 
sequence $\langle c|L\rangle_\mathrm{sat}$ that approaches the red 
sequence with increasing luminosity (equation~\ref{Csatseq}).

\subsection{Central and Satellite Colors as a Function of Luminosity}\label{CcenCsatL}

\begin{figure}
 \includegraphics[width=\hsize]{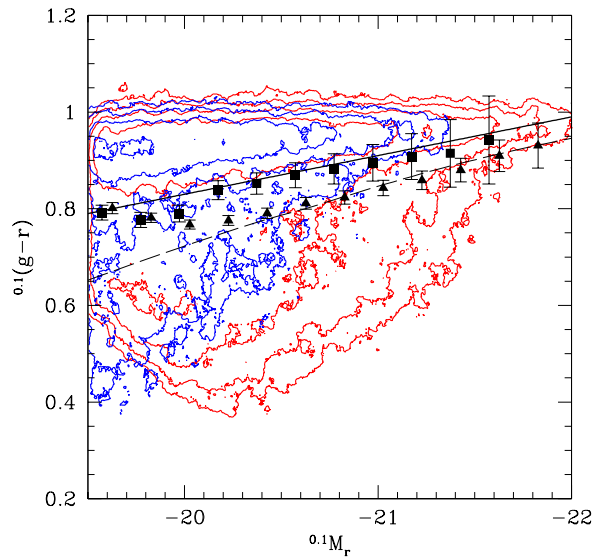} 
 \caption{Color-magnitude diagram of $M_r<-19.5$ volume-limited SDSS catalog
    constructed from the Y07 group catalog.  Contours for central galaxies
    (with brightest $r$-band luminosity) and satellite galaxies are red and blue,
    respectively.  The satellite galaxy color-magnitude sequence 
    $\langle c|L\rangle_\mathrm{sat}$ of the SS08 model (\ref{Csatseq})
    is shown as the thick solid line; the mean satellite galaxy colors at fixed luminosity
    of the group catalog are shown as the square points, with Poisson errors.
    The color-magnitude sequence of central galaxies in the model (\ref{Ccenseq}) 
    and the group catalog are also shown, by the dashed line and triangle points, respectively.
    The measured central and satellite colors (squares and triangles) are slightly offset, for clarity.}
 \label{censatCMD}
\end{figure}

We test the latter assumption first.
In Figure~\ref{censatCMD} we show a color-magnitude diagram with the model's
satellite galaxy color-magnitude sequence.
It is similar to Figure 2 in SS08, but the contours are for a volume-limited 
catalog constructed from the Y07 group catalog, with limits 
$-23.5< {^{0.1}M_r}<-19.5$, $0.017<z<0.082$, consisting of 59,085 galaxies.
The figure shows the color-magnitude contours of central and satellite galaxies 
in the catalog (red and blue contours, respectively).
Central and satellite galaxies appear to have somewhat similar bimodal color
distributions at faint luminosities. 
The majority of bright blue galaxies and luminous red galaxies are
centrals, however, and as we will show later, they tend to reside in less massive 
and more massive halos, respectively.

We compare the satellite galaxy color-magnitude sequence $\langle c|L\rangle_\mathrm{sat}$
(\ref{Csatseq}) of the model to the mean satellite colors in the Y07 group 
catalog in Figure~\ref{censatCMD}.  They are in excellent agreement: 
the model's satellite color sequence is approximately only 0.01 magnitudes 
redder than that of the group catalog and within the Poisson errors across most 
of the luminosity range.
The agreement between the model and the group catalog is encouraging,
and in particular it supports the use of (\ref{Csatseq}) as the satellite 
galaxy color sequence, which could be used as a constraint for 
galaxy formation models that include physical processes that quench the
star formation of satellites.

We also show the color-magnitude sequence of central galaxies in Figure~\ref{censatCMD}.
In the model, this sequence is implied by the satellite sequence (\ref{Csatseq}),
the luminosity-dependent halo occupation distribution, and the observed color-magnitude
constraints:
\begin{equation}
  \langle c|L\rangle_\mathrm{cen} = \langle c|L\rangle_\mathrm{all} + 
 {n_\mathrm{sat}(L)\over n_\mathrm{cen}(L)}\,
 \Bigl[\langle c|L\rangle_\mathrm{all} - \langle c|L\rangle_\mathrm{sat}\Bigr].
 \label{Ccenseq}
\end{equation}
(see SS08 for details).  This sequence is consistent with the measurement from the Y07
catalog, at least for $M_r<-20$.  In both the model and the Y07 catalog, at fixed luminosity,
central galaxies tend to be \textit{bluer} than satellite galaxies.
Van den Bosch et al. (2008a) found the same result, at fixed stellar mass.
However, in a given halo, the central galaxy tends to be significantly brighter and
more massive than its satellites, and it also tends to be redder, as we will show in Section~\ref{CcenCsatN}. 

\begin{figure}
 \includegraphics[width=\hsize]{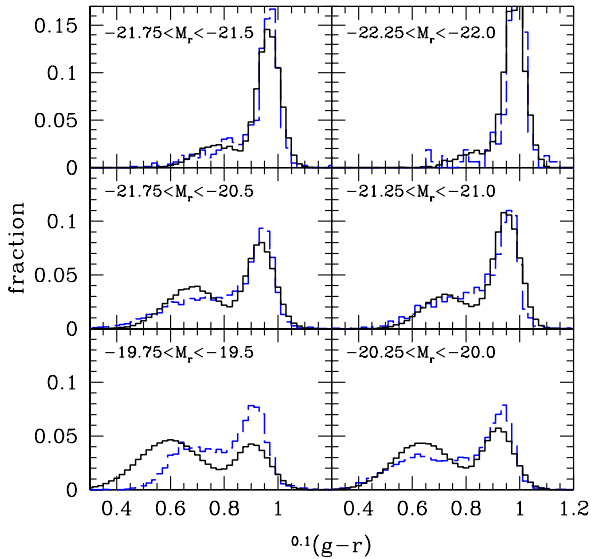} 
 \caption{Distributions of $g-r$ color of central galaxies as a function of luminosity.
          Solid black histograms show the predictions of the model (Skibba \& Sheth 2008);
          Dashed blue histograms show the measurements from the Yang et al. (2007) group
          catalog.}
 \label{CcenLcompare}
\end{figure}

\begin{figure}
 \includegraphics[width=\hsize]{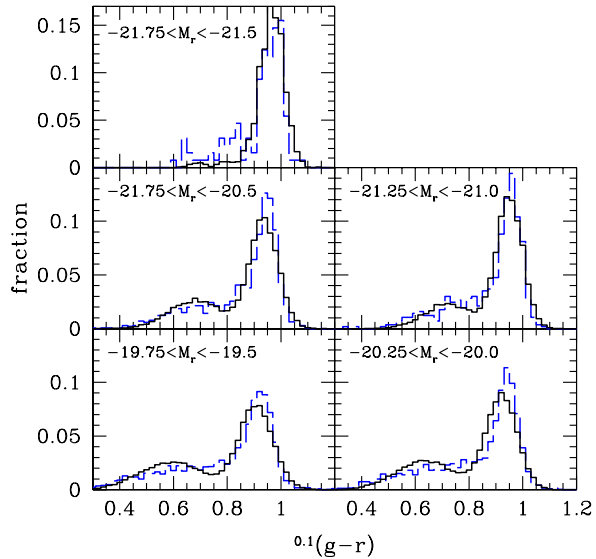} 
 \caption{Same as Figure~\ref{CcenLcompare}, but for satellite galaxies.
          The brightest luminosity bin is excluded because of poor number statistics.}
 \label{CsatLcompare}
\end{figure}

Figure~\ref{censatCMD} showed the mean central and satellite colors as a function of
luminosity, $\langle c|L\rangle_\mathrm{cen}$ and $\langle c|L\rangle_\mathrm{sat}$.
We now compare the central and satellite color distributions, $p_\mathrm{cen}(c|L)$
and $p_\mathrm{sat}(c|L)$, in Figures~\ref{CcenLcompare} and \ref{CsatLcompare}.
The model predictions are determined from a mock galaxy catalog (described in Section~\ref{model}),
and they are compared to measurements from the Y07 group catalog.
%

Overall, the agreement between the model and the group catalog is quite good,
especially for $M_r<-20.5$ (or $L>L_\ast$).
For central galaxies, the bimodality of the color distribution is stronger in the model
than in the data at faint luminosities.  This is partly due to the fact that
the double-Gaussian fit to the color distribution at fixed luminosity is not
perfect, and slightly underpopulates the `green valley' between the red and
blue sequences at faint luminosities.  This does not explain the discrepancy
in the faintest bin, however.
Faint central galaxies tend to reside in low
mass halos or small groups (\textit{e.g.}, Skibba et al. 2007), so the 
discrepancy could be due to inaccuracies in the model or its constraints at low mass
or to inaccuracies in the group catalog in poor groups.  

An interesting result is that both the model and the group catalog agree that the
red sequence and blue sequence are peaked at approximately the same colors at fixed luminosities
for central and satellite galaxies, and that the widths of the red and blue sequences
narrow similarly with increasing luminosity for centrals and satellites.
The difference between the central and satellite color distributions is simply
that the blue fraction at a given luminosity is lower for satellites: 
the red sequence of satellites becomes significantly populated at fainter
luminosity than does the red sequence of centrals.

\subsection{Central and Satellite Colors as a Function of Group Richness}\label{CcenCsatN}

We now investigate the colors of central and satellite galaxies
as a function of group richness, which is strongly correlated with halo mass.
We constructed a volume-limited catalog from the Y07 group catalog
similar to the one above, but with the following limits:
$-23.5<M_r<-19.9$ and $0.015<z<0.100$.
We begin by examining the mean colors of centrals and satellites.
The mean color of central galaxies in groups containing 
$N_\mathrm{gal}$ galaxies was defined as the mean value $\langle g-r\rangle$ 
of all central galaxies of such groups (rather than 
$-2.5\log_{10}\langle L_g/L_r \rangle$ 
or $-2.5\log_{10}\langle L_g\rangle/\langle L_r\rangle$).
The mean satellite color in groups of $N_\mathrm{gal}$ galaxies was computed 
similarly (\textit{i.e.}, the mean of satellite $g-r$).  

We have also measured the mean colors of central and satellite galaxies in the similar
volume-limited catalog of B06.
This $M_r<-19.9$ catalog consists of 21301 galaxies in 4119 groups having 
three or more members.  In comparison, the corresponding Y07 catalog
contains 15234 galaxies in 2163 groups (when fiber-collided galaxies are included), 
which are significantly smaller numbers considering that their catalog was
drawn from a later SDSS data release.
The Berlind et al. catalog has an overabundance of low-$N_\mathrm{gal}$
groups (see their appendix and that of Skibba et al. 2007), which does
not appear to be the case for Yang et al.  However, Berlind et al. only
claim to be complete for $N_\mathrm{gal}\geq10$, and for such richnesses
the two group catalogs are in better agreement.

We will compare the mean central and satellite colors from both group catalogs
to those predicted by the halo occupation model of Skibba \& Sheth (2008).
The model colors are computed the same way as the central and satellite
luminosities were computed in Skibba et al. (2007):
\begin{equation}
 \bigl< c_{\rm cen}|N\bigr> =  \int_{M_{\rm min}(L_{\rm min})}^\infty
    {\rm d}M\, {{\rm d}n(M)\over {\rm d}M}
      {p(N|M)\, \bigl< c_{\rm cen}|M\bigr>
       \over n_{\rm grp}(N)}
 \label{CcenN}
\end{equation}
and 
\begin{equation}
 \bigl< c_{\rm sat}|N,L_{\rm min}\bigr> = 
	 \int_{M_{\rm min}(L_{\rm min})}^\infty {\rm d}M\, 
              {{\rm d}n(M)\over {\rm d}M}\, 
	    {p(N|M)\, \bigl< c_{\rm sat}|M,L_{\rm min}\bigr>
             \over n_{\rm grp}(N)},
 \label{CsatN}
\end{equation}
%
where ${\rm d}n/{\rm d}M$ is the halo mass function, $p(N|M)$ is the halo
occupation distribution, and $n_{\rm grp}(N)$ is the group multiplicity function.

\begin{figure}
 \centering
 \includegraphics[width=\hsize]{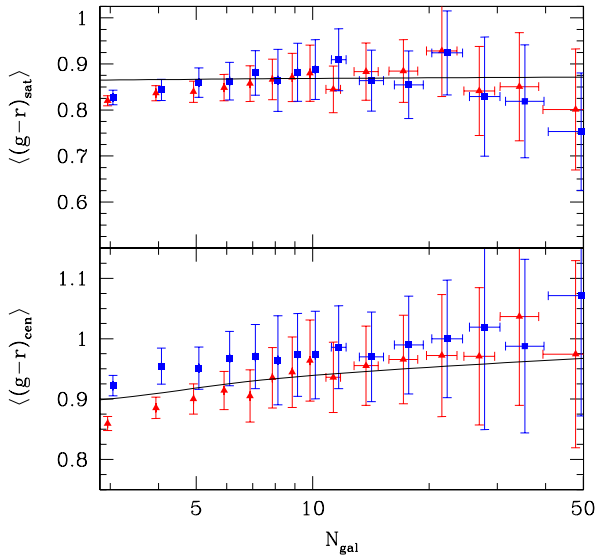} 
 \caption{Mean central and satellite $g-r$ colors as a function of group richness, for $M_r<-19.9$.
          Solid curves show the model predictions (using equations~\ref{CcenN} and \ref{CsatN}),
          and red triangles and blue squares show the measurements from the
          B06 and Y07 galaxy group catalogs, respectively.
          Vertical error bars are the average of the Poisson errors
          (estimated from the number of groups in each bin) and 
          bootstrap errors (estimated from the variance of 10 times as many 
          pseudo-samples as the number of groups).
          Horizontal error bars show the $N_\mathrm{gal}$ bin widths.}
 \label{yangberlind}
\end{figure}

Figure~\ref{yangberlind} shows the results.  
In general, at fixed group richness, satellite galaxies tend to be bluer than
central galaxies, by up to 0.1 mags.  These trends are similar at fixed halo mass 
in the model (not shown).
Because of the dependence of the mean satellite colors on the luminosity threshold
($L_\mathrm{min}$ in eq.~\ref{CsatN}), the difference between the colors of centrals
and satellites would be even larger if fainter galaxies were included.

The model's central and satellite galaxy colors are in very good
agreement with both group catalogs, from poor groups to rich clusters.
The error bars on the measurements are fairly large, but the halo-model 
predictions are also uncertain, due to uncertainties in the luminosity-dependent 
HOD and in the color-magnitude constraints.
The central galaxy colors are systematically redder in the Yang et al. (2007)
catalog than in the Berlind et al. (2006a) catalog.
It could be due to the different cosmology assumed by Yang et al. 
($\Omega_m=0.238$, $\sigma_8=0.75$), or it may be related to the 
different color-dependent clustering at fixed group mass 
measured from the catalogs (Berlind et al. 2006b, Wang et al. 2007).
It is of only weak statistical significance, however.

For the model, we have assumed that the bimodal color distribution at
fixed luminosity is independent of halo mass.
Therefore, the satellite colors are only weakly dependent on
group richness or halo mass because the satellite luminosities are.
Skibba et al. (2007) found that satellite galaxy luminosity is almost
independent of group richness in their halo occupation model and in the 
group catalogs of Yang et al. (2005a) and Berlind et al. (2006a).
The result in the upper panel of Figure~\ref{yangberlind} 
shows that this is also the case for satellite galaxy colors,
and lends justification to the model's assumption that the color distribution at 
given luminosity is independent of mass.
This result has an interesting consequence: the fact that both the luminosity
and color of satellites is nearly independent of halo mass implies
that the evolution of satellite galaxies is nearly independent of the environment.
We discuss this further in Section~\ref{discuss}.

\begin{figure}
 \includegraphics[width=\hsize]{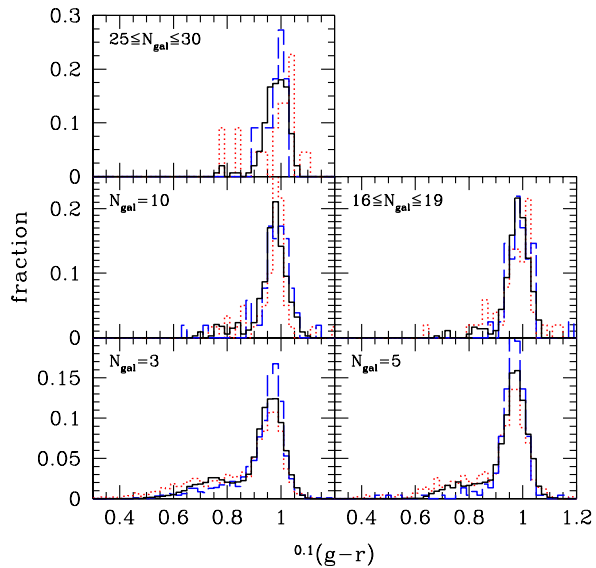} 
 \caption{Distributions of $g-r$ color of central galaxies as a function of group richness,
          for $M_r<-19.9$.  Solid black histograms show the predictions of the model
          (Skibba \& Sheth 2008); dashed blue histograms show the measurements from the
          Yang et al. (2007) group catalog; dotted red histograms show the measurements
          from the Berlind et al. (2006a) catalog.  A very large $N_\mathrm{gal}$ bin
          is not included, because of poor number statistics.}
 \label{CcenNcompare}
\end{figure}

\begin{figure}
 \includegraphics[width=\hsize]{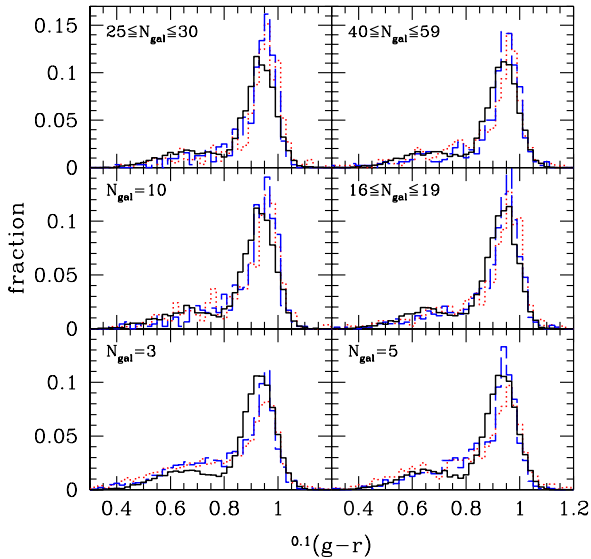} 
 \caption{Same as Figure~\ref{CcenNcompare}, but for satellite galaxies.
          The three large $N_\mathrm{gal}$ bins are chosen to be the same as 
          the $5^\mathrm{th}$-richest, $3^\mathrm{rd}$-richest, and richest 
          bins in Figure~\ref{yangberlind}.}
 \label{CsatNcompare}
\end{figure}

Finally, we compare the central and satellite color distributions as a function of
group richness, $p_\mathrm{cen}(c|N_\mathrm{gal})$ and $p_\mathrm{sat}(c|N_\mathrm{gal})$,
for $M_r<-19.9$, in Figures~\ref{CcenNcompare} and \ref{CsatNcompare}.
The model predictions are determined from a mock galaxy catalog (described 
in Section~\ref{model}), and they are compared to measurements from the 
Yang et al. (2007) and Berlind et al. (2006a) group catalogs.

Overall, there is impressive agreement between the model and group catalogs,
even for groups with few members.
For central galaxies in poor groups, the B06 catalog has a slightly weaker
red peak and a slightly more populated blue bump than in the Y07 catalog,
with the model between them.  For satellite galaxies in poor groups,
the red sequence is peaked at a slightly redder color in the group catalogs
than in the model, and the blue bump is slightly more populated in the catalogs.
These differences are very small, however.

A comparison of the figures makes the following two conclusions evident.
Firstly, for central galaxies, while a significant fraction of them populate
the blue cloud in small groups or low-mass halos, 
the vast majority of them in large groups or massive
halos ($M>10^{14}\,M_\odot/h$) have moved onto the red sequence.
Secondly, for satellite galaxies, the color distribution has a significant
blue bump even in massive groups and clusters.  While there are large numbers
of red satellites in such systems, of course, there are also a significant
number of blue satellites as well, presumably tending to be located in the
cluster outskirts, implied by the color gradients discussed in Section~\ref{model}.
Moreover, the satellite color distribution is almost independent of group
richness or halo mass.  This occurs in the model by construction: we have assumed 
that the color distribution at fixed luminosity is independent of mass, and 
satellite luminosity only weakly depends on richness.  The agreement with the
group catalogs constitutes evidence in support of this assumption.

\section{Discussion}\label{discuss}

To summarize, we have compared the halo model of galaxy colors of Skibba \& 
Sheth (2008) to the colors of galaxies in the Yang et al. (2007) and Berlind 
et al. (2006a) group catalogs.
The model assumes that satellite galaxies tend to follow a particular sequence
along the color-magnitude diagram (\ref{Csatseq}), such that it approaches the 
red sequence at bright luminosities, and we have found that this satellite 
color sequence is in excellent agreement with measurements from the Yang et al.
group catalog.
This constitutes support for the Skibba \& Sheth model, and for the satellite
color sequence itself, which could be used as a constraint for galaxy formation
models on the physical processes that quench the star formation of satellite 
galaxies. 
The satellite color sequence as a function of luminosity can easily be 
converted into a sequence as a function of halo mass as well (see Section 2.2
of Skibba \& Sheth).
The agreement between the model and the group catalog suggests that a 
significant fraction of faint satellites (with $L<L_\ast$), which reside
in low-mass halos as well as cluster-sized halos (Skibba et al. 2007),
are still forming stars out of a dwindling gas supply and are in the
process of being quenched and transformed into `red and dead' galaxies.

Van den Bosch et al. (2008a) recently used the Yang et al (2007) group catalog 
to explore the impact of various transformation mechanisms that are believed to 
operate on satellite galaxies.  Based on the colors and concentrations of 
galaxies at fixed stellar mass, they found that the main mechanism that causes 
the transition of satellite galaxies from the blue to the red sequence is 
strangulation, in which the hot diffuse gas around newly accreted satellites is 
stripped, removing its fuel for future star formation.
They ruled out other mechanisms, such as ram-pressure stripping, which is efficient
only in massive halos, and harassment, which alters galaxies' morphologies.
Kang \& van den Bosch (2008) used these results with their semi-analytic model
to show that strangulation is inefficient and takes a few Gyr to operate.
A similar conclusion was recently obtained by Font et al. (2008). who used
a more sophisticated stripping model.
Cattaneo et al. (2008) have recently used another semi-analytic model to show 
that the observed `archaeological downsizing', in which stars in more massive 
galaxies tend to form earlier and over a shorter period, 
can be reproduced if strangulation shuts down star formation only above
a critical halo mass $M_\mathrm{crit}\sim10^{12}M_\odot$.
Gilbank \& Balogh (2008) came to the same conclusion when attempting to
reproduce the red sequence dwarf-to-giant ratio in clusters and the field.
This critical mass is slightly larger than the minimum halo mass for $M_r<-19.5$,
and at this mass scale the satellite color sequence of the Skibba \& Sheth (2008) 
model is indeed approaching the red sequence.

We also analyzed the colors of central and satellite galaxies as a
function of group richness, quantified by the number of galaxies in a group
more luminous than a given threshold.
We showed that at fixed richness or halo mass, central galaxies tend to be redder
than satellites, and the color difference increases with richness.
In contrast, at fixed luminosity or stellar mass, centrals tend to be bluer
than satellites.
This is simply explained by the fact that in a given halo, the central galaxy is 
usually the brightest and most massive galaxy.
These results suggest that as central and satellite galaxies evolve, they may follow 
different paths along the color-magnitude diagram.
For example, central galaxies become more luminous as they evolve,
whether secularly or with mergers, 
and may continue to form stars and remain blue even after forming a significant bulge
component, and then move towards the red sequence after AGN feedback has operated; 
on the other hand, satellites may experience star formation quenching 
due to strangulation and approach the red sequence while they are still faint and disk-dominated.
These issues are investigated further in Skibba et al. (2008a, in prep.),
using morphology mark correlation functions with the SDSS Galaxy Zoo 
catalog of visually classified morphologies.

For the colors of both central and satellite galaxies as a function of group richness, 
the model of Skibba \& Sheth (2008) and the two group catalogs of 
Yang et al. (2007) and Berlind et al. (2006a) are in very good agreement.
Central galaxies tend to be the reddest galaxy in a halo and are much redder
than the typical satellite in large groups.  Satellite galaxy color, unlike that of centrals,
is almost independent of group richness.
The distributions of central and satellite colors as a function of richness
in the model and group catalogs are also in good agreement.  Central galaxies
have a bimodal color distribution in small groups, but the vast majority of them
in larger groups have moved onto the red sequence.  In contrast,
the satellite color distribution is almost independent of group richness.
This occurs in the model by construction: we assumed that the
color distribution at fixed luminosity is independent of halo mass,
and satellite luminosity only weakly depends on richness. 
The agreement with the group catalogs supports this assumption.

It is worth emphasizing that, while some authors focus on mean
galaxy properties (\textit{e.g.}, Mart\'{i}nez \& Muriel 2006, Conroy \& Wechsler 2008)
or on red or blue fractions (\textit{e.g.}, Weinmann et al. 2006, van den Bosch et al. 2008a),
our model has predicted the mean and \textit{distributions} of 
central and satellite galaxy colors, as a function of luminosity
and richness, in agreement with the SDSS data for $M_r<-20$.
This is quite a feat, considering that our model is fairly simple,
based on luminosity-dependent clustering and color-magnitude constraints,
with very few assumptions.

It is interesting that satellite galaxy color appears to be almost independent
of host halo mass, 
and that this is also the case for satellite galaxy luminosity (Skibba et al. 2007).
Since galaxy color is tightly correlated with stellar mass-to-light ratio 
(Bell et al. 2003), this implies that satellite galaxies of a given 
stellar mass can also be found in halos of a wide range of masses.
This suggests that what most determines a satellite galaxy's properties, 
and its evolution in general, is its stellar mass, not its host halo mass.
In addition, galaxy color and luminosity are the primary properties that are
most predictive of a galaxy's environment (Blanton et al. 2005a), so the
flat relations of satellite galaxy color and luminosity with halo mass suggest a dearth
of environmental dependence for the transformation of satellite galaxies.
This point was recently made by van den Bosch et al. (2008b), who used the
Yang et al. (2007) group catalog to show that the color and concentration
of satellite galaxies are almost completely determined by their stellar
mass, with only a very weak dependence on halo mass and halo-centric radius.
One consequence of this is that `pre-processing' in groups cannot be the
dominant process that differentiates the cluster galaxy population from
that of the field.

Finally, Brown et al. (2008) recently completed a study of the evolution of 
the luminosity and stellar mass of red central and satellite galaxies, 
using halo occupation models.
We can do a few simple comparisons between our results and theirs.
They find that the stellar masses of luminous red central galaxies scales 
with halo mass to the power of $\approx0.35$.  Using our model's 
relationship between central galaxy color and halo mass (lower panel of 
Figure~\ref{yangberlind}) with the relation between colors and stellar 
mass-to-light ratios (Bell et al. 2003), we can estimate the relation 
between central galaxy stellar mass and halo mass for the model.  We find 
that the slope of this relation approaches $\approx0.37$, similar to their 
result.  Brown et al. also show that approximately 50\% of 
$10^{11.9}\,h^{-1}\,M_\odot$ mass halos host central galaxies that are red,
and this fraction increases with halo mass.  
Our model predicts a fraction of $\approx45\%$ at this mass, 
although our separation of `red' and `blue' galaxies is in terms of $g-r$ 
color, while theirs is in terms of $U-V$.
Finally, Brown et al. also conclude that the fraction of stellar mass within
the satellite population increases with host halo mass, which is consistent
with our results and Skibba et al. (2007).


\section*{Acknowledgements}
I would like to thank Xi Kang, Frank van den Bosch, 
and Ravi Sheth for valuable discussions, and I thank
Frank van den Bosch for providing the galaxy group 
catalog of Yang et al (2007).

Funding for the SDSS and SDSS-II has been provided by the 
Alfred P. Sloan Foundation, the Participating Institutions, 
the National Science Foundation, the U.S. Department of Energy, 
the National Aeronautics and Space Administration, 
the Japanese Monbukagakusho, the Max Planck Society, 
and the Higher Education Funding Council for England. 
The SDSS Web Site is http://www.sdss.org/.

The SDSS is managed by the Astrophysical Research Consortium for 
the Participating Institutions. The Participating Institutions are 
the American Museum of Natural History, Astrophysical Institute 
Potsdam, University of Basel, Cambridge University, 
Case Western Reserve University, University of Chicago, 
Drexel University, Fermilab, the Institute for Advanced Study, 
the Japan Participation Group, Johns Hopkins University, 
the Joint Institute for Nuclear Astrophysics, 
the Kavli Institute for Particle Astrophysics and Cosmology, 
the Korean Scientist Group, the Chinese Academy of Sciences (LAMOST), 
Los Alamos National Laboratory, 
the Max-Planck-Institute for Astronomy (MPA), 
the Max-Planck-Institute for Astrophysics (MPIA), 
New Mexico State University, Ohio State University, 
University of Pittsburgh, University of Portsmouth, 
Princeton University, the United States Naval Observatory, 
and the University of Washington.

\label{lastpage}

\end{document}